\documentclass[twocolumn,preprintnumbers,amsmath,amssymb,aps,superscriptaddress]{revtex4}

\setcitestyle{super}

\usepackage{graphicx}
\usepackage{dcolumn}
\usepackage{bm}
\begin{document}

\title{Josephson parametric phase-locked oscillator and its
application to dispersive readout of superconducting qubits}

\author{Z. R. Lin} 
\affiliation{RIKEN Center for Emergent Matter Science (CEMS), Wako, Saitama 351-0198, Japan}

\author{K. Inomata} 
\affiliation{RIKEN Center for Emergent Matter Science (CEMS), Wako, Saitama 351-0198, Japan}

\author{K. Koshino}
\affiliation{College of Liberal Arts and Sciences, Tokyo Medical and Dental University, 
Ichikawa, Chiba 272-0827, Japan}

\author{W. D. Oliver}
\affiliation{MIT Lincoln Laboratory, Lexington, Massachusetts 02420, USA}

\author{Y. Nakamura}
\affiliation{Research Center for Advanced Science and Technology (RCAST), 
The University of Tokyo, Meguro-ku, Tokyo 153-8904, Japan}
\affiliation{RIKEN Center for Emergent Matter Science (CEMS), Wako, Saitama 351-0198, Japan}
 
\author{J. S. Tsai}
\affiliation{RIKEN Center for Emergent Matter Science (CEMS), Wako, Saitama 351-0198, Japan}
\affiliation{NEC Smart Energy Research Laboratories, Tsukuba, Ibaraki 305-8501, Japan}

\author{T. Yamamoto} \email[t-yamamoto@fe.jp.nec.com]{}
\affiliation{RIKEN Center for Emergent Matter Science (CEMS), Wako, Saitama 351-0198, Japan}
\affiliation{NEC Smart Energy Research Laboratories, Tsukuba, Ibaraki 305-8501, Japan}

\date{\today}

\maketitle

{\bf 
The parametric phase-locked oscillator (PPLO),~\cite{Onyshkevych59} 
also known as a parametron,~\cite{Goto59} is a resonant circuit 
in which one of the reactances is periodically modulated. 
It can detect, amplify, and store binary digital signals 
in the form of two distinct phases of self-oscillation. 
Indeed, digital computers using PPLOs based on a magnetic ferrite ring 
or a varactor diode as its fundamental logic element 
were successfully operated in 1950s and 1960s.~\cite{Goto59}
More recently, basic bit operations have been demonstrated 
in an electromechanical resonator,~\cite{Mahboob08} 
and an Ising machine based on optical PPLOs has been proposed.~\cite{Wang13}
Here, using a PPLO realized with Josephson-junction circuitry, 
we demonstrate the demodulation of a microwave signal 
digitally modulated by binary phase-shift keying. 
Moreover, we apply this demodulation capability 
to the dispersive readout of a superconducting qubit. 
This readout scheme enables a fast and latching-type readout, yet 
requires only a small number of readout photons in the resonator to which the qubit is coupled, 
thus featuring the combined advantages of several disparate schemes.~\cite{Siddiqi06,Vijay11} 
We have achieved high-fidelity, single-shot, and non-destructive qubit readout 
with Rabi-oscillation contrast exceeding 90\%, limited primarily by 
the qubit's energy relaxation. 
}

\begin{figure}
\begin{center}
\includegraphics[width=3.2 in]{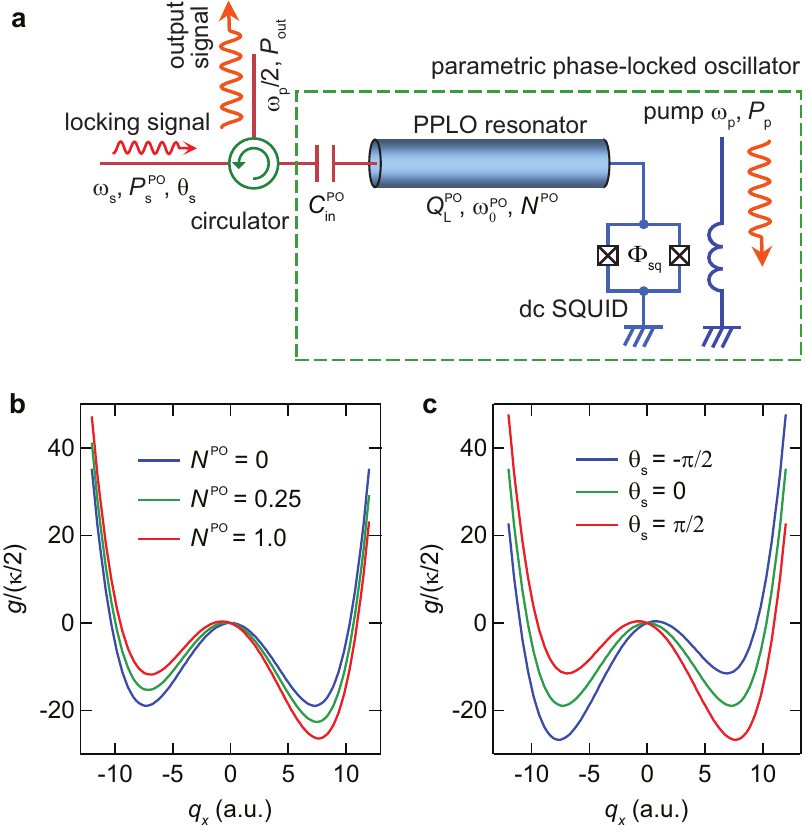}
\end{center}
\caption{\label{fig1}{\bf Parametric phase-locked oscillator.}
{\bf a,} Schematic of the device. 
The application of a pump microwave field at $\omega_{\rm p} \sim 2\omega_0^{\rm PO}$ 
with a power $P_{\rm p}$ above the threshold generates an 
output signal at $\omega_{\rm p}/2$ with a power $P_{\rm out}$. 
Here, $\omega_0^{\rm PO}$ is the static resonant frequency of the PPLO resonator, and 
$\omega_{\rm s}$, $P_{\rm s}^{\rm PO}$ and $\theta_{\rm s}$ represent 
the frequency, power and phase of the locking signal, respectively. 
$C_{\rm in}^{\rm PO}$ represents the coupling capacitance between the resonator 
and the feedline, and $\Phi_{\rm sq}$ represents the magnetic flux 
penetrating the SQUID loop. 
{\bf b,} Cross-section of the Hamiltonian $g(q_x,0)$ 
in the rotating frame for various powers and a fixed phase ($\theta_{\rm s}=\pi/2$) 
of the locking signal. 
Here, the LS power is represented by the mean photon number in the PPLO resonator, 
$N^{\rm PO}
=4P_{\rm s}^{\rm PO}{Q_{\rm L}^{\rm PO}}^2/(\hbar{\omega_0^{\rm PO}}^2Q_{\rm e}^{\rm PO})$. 
The pump power $P_{\rm p}/P_{\rm p0}$ is fixed at 2.0. 
Two minima correspond to $0\pi$ and $1\pi$ states of the phase-locked oscillations. 
{\bf c,} Same as {\bf b}, but for $N^{\rm PO}=1$ 
and various values of $\theta_{\rm s}$. 
}
\end{figure}

Our PPLO is implemented by a superconducting coplanar waveguide (CPW) resonator 
defined by a coupling capacitor $C_{\rm in}^{\rm PO}$ and a dc-SQUID termination 
(Fig.~\ref{fig1}a; see Supplementary Information for the details of the device). 
The dc-SQUID, working as a controllable inductor, 
makes the resonant frequency of the resonator $\omega_0^{\rm PO}$ dependent on 
an external magnetic flux through the SQUID loop $\Phi_{\rm sq}$~(Ref.~\citenum{Wallquist06}). 
Thus, the application of a microwave field at a frequency $\omega_{\rm p}$ and a power 
$P_{\rm p}$ to the pump line, which is inductively coupled to the SQUID loop, 
modulates the resonant frequency around its static value. 
This device has previously been operated 
as a Jopsehson parametric amplifier~\cite{Yamamoto08,Lin13} (JPA): 
The signal at $\sim \omega_0^{\rm PO}$ entering the resonator obtains a parametric gain 
produced by the pump at $\omega_{\rm p} \sim 2\omega_0^{\rm PO}$ and 
is reflected back along the signal line. 
This device also works as a parametric oscillator when it is operated at $P_{\rm p}$ 
above the threshold $P_{\rm p0}$ determined 
by the photon decay rate of the resonator.~\cite{Wilson10,Krantz13}  
Namely, it generates an output microwave field at $\omega_{\rm p}/2$ 
even without any signal injection. The output can be one of the 
two degenerate oscillatory states ($0\pi$ and $1\pi$ states) 
which differ solely by a relative phase shift $\pi$. 

When an additional signal at $\sim \omega_{\rm p}/2$ with a power $P_{\rm s}^{\rm PO}$, 
which we call the locking signal (LS), 
is injected into the parametric oscillator, 
the degeneracy of the two oscillatory states is lifted.~\cite{Ryvkine06} 
Such degeneracy lifting has been demonstrated in other physical systems 
such as magneto-optically trapped cold atoms~\cite{Kim10} 
and electromechanical systems.~\cite{Mahboob10}
To understand the dynamics, we consider the Hamiltonian of the PPLO 
including a signal port for the LS and 
a fictitious loss port for internal loss of the resonator, namely, 
\begin{eqnarray} \label{hamiltonian}
&&\mathcal{H}/\hbar = \omega_0^{\rm PO} a^\dagger a + 
\frac{\kappa}{2}\sqrt{\frac{P_{\rm p}}{P_{\rm p0}}} \cos(2\omega_0^{\rm PO} t) (a + a^\dagger)^2 
\nonumber \\
&& + \gamma (a + a^\dagger)^4 + \int dk \Bigl[ v_bk b_k^\dagger b_k + \mathrm{i} \sqrt{\frac{v_b\kappa_1}{2 \pi}}
\Bigl( a^\dagger b_k-b_k^\dagger a \Bigr) \Bigr] \nonumber \\
&& + \int dk \Bigl[ v_ck c_k^\dagger c_k + \mathrm{i} \sqrt{\frac{v_c\kappa_2}{2 \pi}}
\Bigl( a^\dagger c_k-c_k^\dagger a \Bigr) \Bigr], 
\end{eqnarray}
where $\gamma$ represents the nonlinearity of the Josephson junction (JJ), 
$a$ is the annihilation operator for the resonator, 
$b_k$ ($c_k$) is the annihilation operator for the photon in the signal (loss) port 
with a wave number $k$ and a velocity $v_b$ ($v_c$), 
$\kappa_1$ ($\kappa_2$) represents the coupling strength between the resonator 
and the signal (loss) port, and $\kappa=\kappa_1+\kappa_2$. 
The equations of motion for the classical resonator field in a frame rotating at 
$\omega_0^{\rm PO}$ are obtained from the Hamiltonian 
by setting $\langle a \rangle = [q_x(t)-\mathrm{i}q_y(t)]e^{-\mathrm{i}\omega_0^{\rm PO}t}$ 
and are given by (Ref.~\citenum{Dykman98}; also see Supplementary Information for derivation) 
\begin{eqnarray}
\frac{dq_x}{dt} &=& -\frac{\kappa}{2}q_x + \frac{\partial g}{\partial q_y}, \label{eqm_qx}\\
\frac{dq_y}{dt} &=& -\frac{\kappa}{2}q_y - \frac{\partial g}{\partial q_x}, \label{eqm_qy}
\end{eqnarray}
where 
\begin{eqnarray}
g(q_x,q_y) = &&\frac{\kappa}{4} \sqrt{\frac{P_{\rm p}}{P_{\rm p0}}}(q_y^2-q_x^2) 
- 3\gamma(q_x^2+q_y^2)^2 
\nonumber \\ &&+ \sqrt{\kappa_1}|E_{\rm s}|(q_y\cos\theta_{\rm s} - q_x\sin\theta_{\rm s}), 
\end{eqnarray}
and $|E_{\rm s}|=\sqrt{P_{\rm s}^{\rm PO}/\hbar\omega_0^{\rm PO}}$ 
and $\theta_{\rm s}$ are the amplitude 
and phase of the LS, respectively. 
In the absence of the first term (oscillator's friction term) on the right-hand side, 
Eqs.~\ref{eqm_qx} and \ref{eqm_qy} are in the form of Hamilton's equations of motion 
with the Hamiltonian $g(q_x,q_y)$, whose minima correspond to 
$0\pi$ and $1\pi$ states.~\cite{Dykman98} 
As shown in Fig.~\ref{fig1}b, $g(q_x,0)$ is symmetric with respect to $q_x$ 
when there is no LS ($|E_{\rm s}|=0$), and $0\pi$ and $1\pi$ states are degenerate. 
When we apply LS, it gives a tilt to the double well, 
which is proportional to $|E_{\rm s}|\sin\theta_{\rm s}$ 
(Figs.~\ref{fig1}b and~\ref{fig1}c). 
This lifts the degeneracy, and the PPLO, initially at $q_x,q_y \sim 0$, preferably 
evolves into one of the two states. 
This is how the amplitude and the phase of LS control the 
output state of the PPLO. 

\begin{figure*}
\begin{center}
\includegraphics[width=5.95 in]{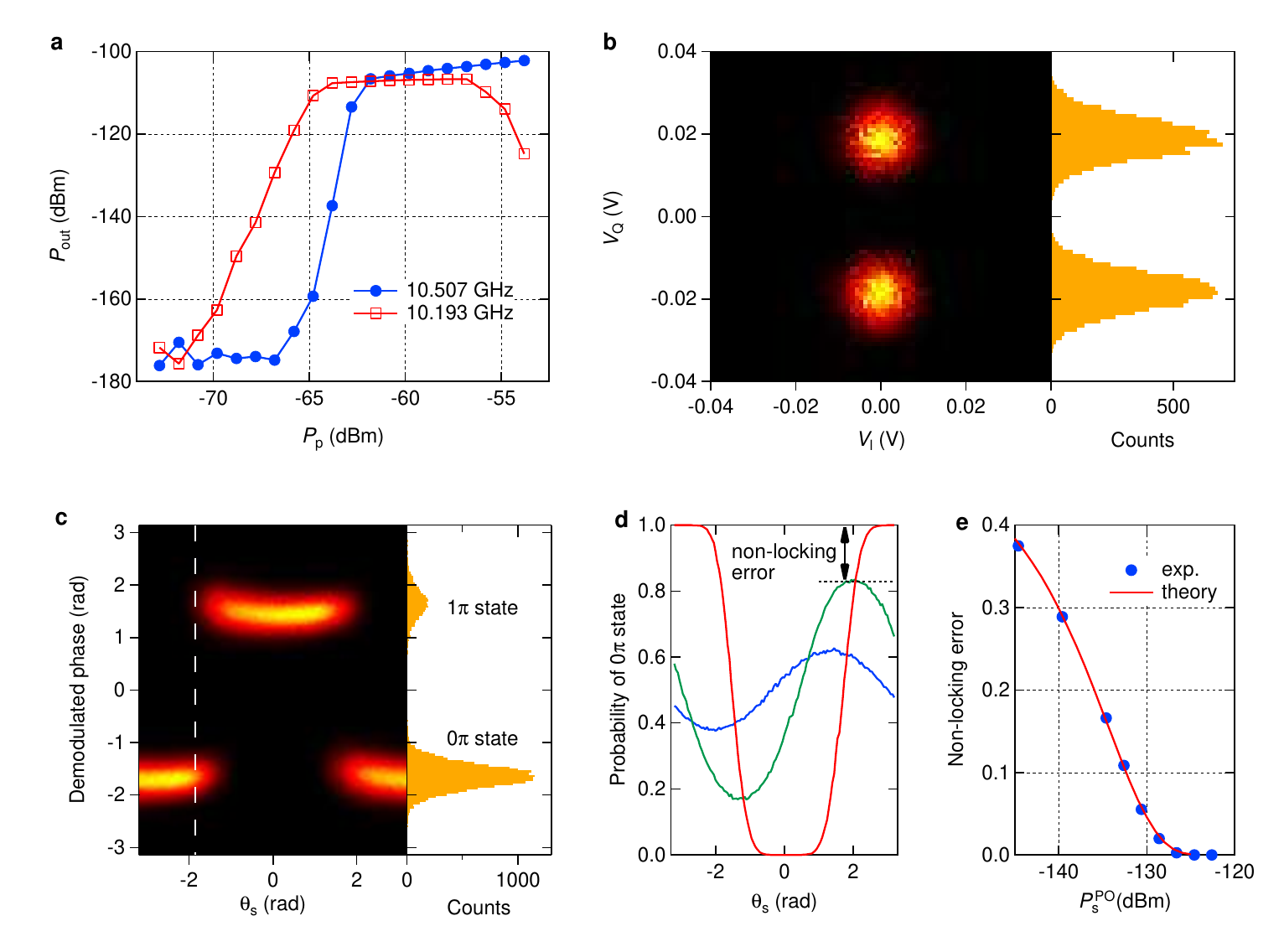}
\end{center}
\caption{{\bf Output of the parametric phase-locked oscillator (PPLO).}
{\bf a,} Output signal power as a function of $P_{\rm p}$ at 
$\omega_{\rm 0}^{\rm PO}/2\pi=10.507$~GHz (solid blue circles) 
and $\omega_{\rm 0}^{\rm PO}/2\pi=10.193$~GHz (open red squares). 
The steep increase of the output power indicates onset of the parametric oscillation. 
For {\bf b}-{\bf e}, $\omega_{\rm 0}^{\rm PO}/2\pi$ is set to be 10.507~GHz. 
{\bf b,} Histogram of the amplitude and phase of the 
output signal plotted in the quadrature (IQ) plane (left panel), 
and its projection onto the $V_{\rm Q}$ axis (right panel). 
The pulsed pump with a duration of $1.6~\mu$s and an amplitude of $-62$~dBm 
is applied 1.6$\times10^4$ times in total. No locking signal (LS) is applied. 
{\bf c,} Histogram of the output-signal phase 
as a function of the LS phase $\theta_{\rm s}$ (left panel) 
and its cross-section along the dashed line (right panel). 
The same pulsed pump as in {\bf b} 
is applied 2.0$\times10^4$ times for each $\theta_{\rm s}$. 
LS is continuously applied with the power of $-125$~dBm. 
{\bf d,} Probability of $0\pi$ state as a function of 
$\theta_{\rm s}$ for $P_{\rm s}$ of $-145$~dBm (blue), $-135$~dBm (green), and $-125$~dBm (red). 
{\bf e,} Non-locking error as a function of the LS power $P_{\rm s}^{\rm PO}$. 
Solid blue circles represent the experimental data, and the red curve is 
the result of a simulation based on the master equation. 
~\label{fig2}
}
\end{figure*}

Now we show the experimental results. 
Figure~\ref{fig2}a shows the output power of the PPLO operated 
at $\omega_{\rm p}=2\omega_0^{\rm PO}=2\pi\times21.014$~GHz (solid blue circles) 
as a function of $P_{\rm p}$ in the absence of LS injection. 
For each $P_{\rm p}$, the microwave field is continuously applied to the pump port, 
and the output signal at $\omega_{\rm p}/2$ is measured by a spectrum analyzer. 
Note that power levels stated in this work are referred to the corresponding ports on the chip. 
The steep increase of the output power at $P_{\rm p} \sim -65$~dBm indicates the onset 
of the parametric oscillation. 
This is further confirmed by detecting the response to the pulsed pump. 
Figure~\ref{fig2}b shows the histogram of the demodulated amplitude and 
phase of the output signal plotted in the quadrature (IQ) plane. 
For each application of the pump pulse with an amplitude of $P_{\rm p}=-62$~dBm, 
we recorded the output pulse and extracted its amplitude and phase by averaging for 100~ns. 
The two distribution peaks correspond to $0\pi$ and $1\pi$ states. 
They have equal amplitude, but different phases shifted by $\pi$, 
and are observed with equal probabilities as expected. 

Next we perform a similar measurement, but include LS injection. 
Figure~\ref{fig2}c shows the histogram of the demodulated phase of the output signal 
as a function of the LS phase.
While continuously injecting LS with a phase $\theta_{\rm s}$ and a power 
$P_{\rm s}^{\rm PO} = -125$~dBm, we applied a pulsed pump 
with the same duration and the amplitude as in Fig.~\ref{fig2}b, 
and extracted the phase of the output signal. 
As exemplified by the cross-section along the dashed line, 
the probabilities of obtaining the two states are no longer 
equal and depend on $\theta_{\rm s}$. 
In Fig.~\ref{fig2}d, we plot the probability of $0\pi$ state 
as a function of $\theta_{\rm s}$ for different $P_{\rm s}^{\rm PO}$'s. 
The probability shows sinusoidal dependence when $P_{\rm s}^{\rm PO}$ is small. 
As we increase $P_{\rm s}^{\rm PO}$, the modulation amplitude 
($\Delta P_{0\pi}$) also increases, and finally reaches unity. 
We define $(1-\Delta P_{0\pi})/2$ as a non-locking error, 
and plot it as a function of $P_{\rm s}^{\rm PO}$ 
in Fig.~\ref{fig2}e by blue circles. 
When $P_{\rm s}^{\rm PO}$ is larger than $\sim-125$~dBm, 
the non-locking error becomes negligible. 
We also simulated the non-locking error by solving a master equation~\cite{Johansson13} 
based on the Hamiltonian (Eq.~\ref{hamiltonian}), and plot it by red curve. 
(For details, see Supplementary Information.) 
The only assumption here is the pump threshold $P_{\rm p0}$, 
which is set to be $-64.0$~dBm. 
The agreement between the theory and the experiment is fairly good.

\begin{figure}
\begin{center}
\includegraphics[width=3.35 in]{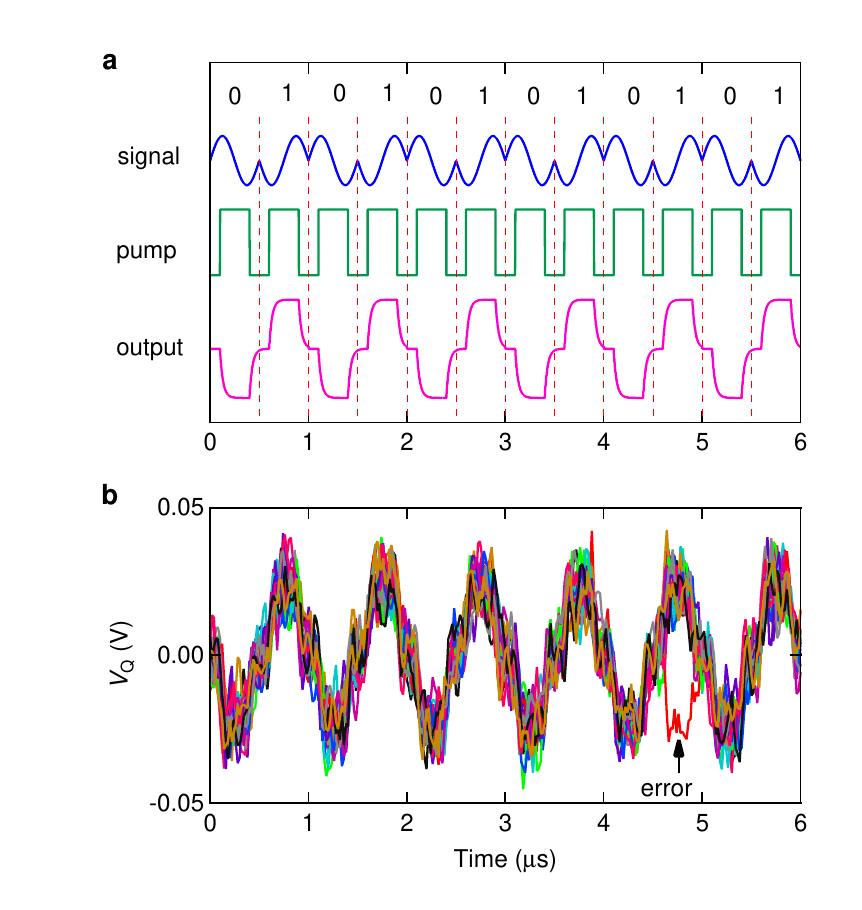}
\end{center}
\caption{{\bf Demodulation of BPSK signal.} 
{\bf a,} Pulse sequence. 
BPSK signal at 10.503~GHz with an amplitude of $-125$~dBm 
is generated by mixing the 10.553~GHz signal and 50~MHz IF signal (blue curve). 
Phase of the IF signal is digitally modulated by $\pi$ (corresponding to 
bit ``0'' and bit ``1'') every 500~ns, 
and the pulsed pump (green curve) is synchronously applied. 
The magenta curve represents the expected output of the PPLO. 
{\bf b,} Examples of the demodulated signal. 
Twenty individual time traces are superposed. 
~\label{fig3}
}
\end{figure}

The above result indicates the PPLO is a phase detector sensitive to 
very small microwave powers of the order of a femtowatt. 
To demonstrate this, we generate a signal digitally modulated 
by binary phase-shift keying (BPSK), a scheme commonly used in modern 
telecommunications,~\cite{AndersonBook} and demodulate it using a PPLO. 
Figure~\ref{fig3}a shows the sequence of the experiment. 
The generated signal has a fixed carrier frequency of 10.503~GHz, fixed power of $-125$~dBm, 
and a phase which is digitally modulated by $\pi$ every 500~ns. 
This means that the signal carries alternating binary bits 
with a baseband frequency of 2~MHz. 
Synchronously, we apply pump pulses with the duration of 300~ns and 
the amplitude of $-62$~dBm. 
Figure~\ref{fig3}b shows the superposed time traces of the PPLO output.  
It shows successful demodulation of the input signal 
except rare errors as exemplified in the figure. 
We sent a total of $2.4\times10^4$ bits and 
detected 4 errors, corresponding to the error rate of $1.7\times10^{-4}$.

\begin{figure*}
\begin{center}
\includegraphics[width=5.94 in]{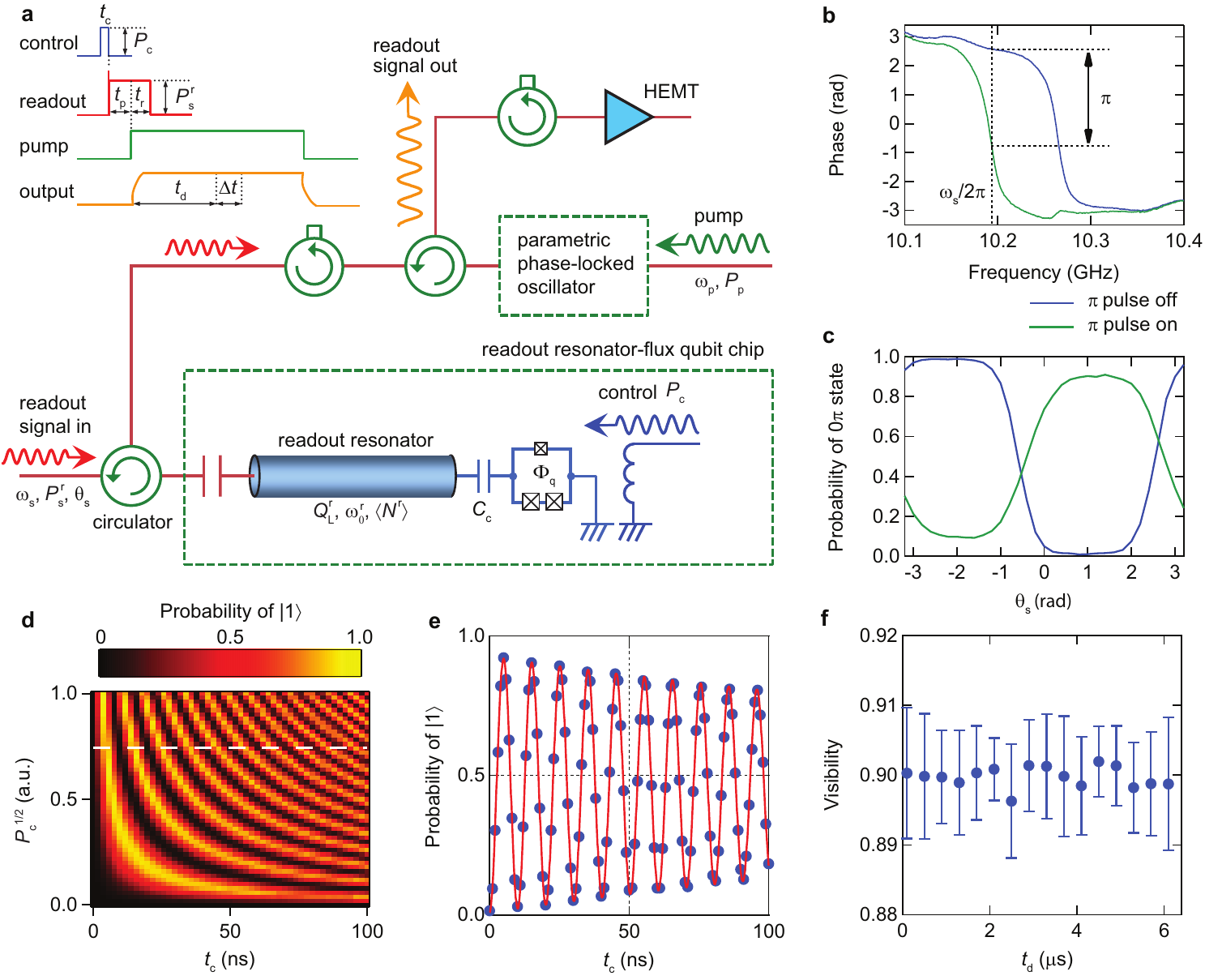}
\end{center}
\caption{{\bf Qubit readout using PPLO.} 
{\bf a,} Measurement setup and pulse sequence. 
A 3-JJ superconducting flux qubit biased at $\Phi_{\rm q}=\Phi_0/2$ 
is capacitively ($C_{\rm c}=4$~fF) coupled to a CPW resonator. 
Here, $\omega_0^{\rm r}$ and $N^{\rm r}$ represent the resonant frequency 
and the mean photon number of the readout resonator, respectively. 
The internal ($Q_{\rm i}^{\rm r}$) and external ($Q_{\rm e}^{\rm r}$) quality factors 
are $2.5\times 10^4$ and 630, respectively, 
resulting in the loaded quality factor $Q_{\rm L}^{\rm r}$ of 620. 
In the pulse sequence, $\Delta t$~(=100~ns) represents the data-acquisition 
time to extract the phase, and $P_{\rm s}^{\rm r}$ and $P_{\rm c}$ represent 
the power of the readout and the qubit-control pulses, respectively. 
A short spike was added in the readout pulse to improve the readout fidelity. 
In {\bf c} to {\bf f}, we applied the pulse sequence $1\times10^4$ times to 
obtain the probability. 
{\bf b,} Frequency dependence of the phase of the reflection coefficient of the readout resonator 
with qubit-control $\pi$ pulse on (green) and off (blue). 
{\bf c,} Probability of the PPLO $0\pi$ state as a function of the readout-signal phase 
$\theta_{\rm s}$ with the qubit-control $\pi$ pulse on (green) and off (blue). 
Here, $t_{\rm c}=10$~ns, $t_{\rm p}=50$~ns, $t_{\rm r}=50$~ns, and $t_{\rm d}=300$~ns. 
{\bf d,} Rabi oscillations measured with 
$t_{\rm p}=30$~ns, $t_{\rm r}=50$~ns, and $t_{\rm d}=300$~ns. 
{\bf e,} Rabi oscillations along the dashed line in {\bf d}. 
Red curve represents a fit with an exponentially damped sinusoidal function. 
{\bf f,} Visibility as a function of $t_{\rm d}$. 
Here, $t_{\rm c}=10$~ns, $t_{\rm p}=30$~ns, and $t_{\rm r}=50$~ns. 
The error bars represents the standard deviation in five identical measurements. 
~\label{fig4}
}
\end{figure*}

Now we apply this phase discrimination capability to the dispersive readout of a qubit. 
Figure~\ref{fig4}a shows the measurement setup. 
A chip containing a 3-JJ superconducting flux qubit 
capacitively coupled to a CPW resonator 
(readout resonator) is connected to the PPLO via circulators. 
The flux qubit is biased at $\Phi_{\rm q}=0.5\Phi_0$, 
where the qubit transition frequency from $|0\rangle$ to $|1\rangle$ states is 5.510~GHz. 
Figure~\ref{fig4}b shows the phase rotation of the reflection coefficient 
around the resonant frequency of the readout resonator ($\omega_0^{\rm r}$) 
when the qubit is in $|0\rangle$ (blue) 
and $|1\rangle$ (green). 
The shift between the two curves is due to the dispersive coupling 
between the qubit and the readout resonator.~\cite{Inomata12} 
We set $\omega_{\rm s}$ to be 10.193~GHz, 
such that the qubit states are mapped onto two phases 
of the reflected microwave differing by $\pi$. 

We discriminate the two states by using the PPLO and 
the pulse sequence shown in Fig.~\ref{fig4}a. 
We set $\omega_{\rm 0}^{\rm PO}=\omega_{\rm s}$, $\omega_{\rm p}=2\omega_{\rm s}$ 
(corresponding to open red squares in Fig.~\ref{fig2}a), and $P_{\rm p}=-65$~dBm. 
Figure~\ref{fig4}c shows the probability of detecting the $0\pi$ state 
as a function of the phase of the readout microwave pulse $\theta_{\rm s}$
(similar to Fig.~\ref{fig2}d) with the qubit control 
$\pi$ pulse off (blue) and on (green). 
The power of the readout microwave field at the input of the readout resonator 
$P_{\rm s}^{\rm r}$ is $-120$~dBm, which corresponds to the mean photon number 
in the readout resonator $N^{\rm r}= 
4P_{\rm s}^{\rm r}{Q_{\rm L}^{\rm r}}^2/(\hbar{\omega_0^{\rm r}}^2Q_{\rm e}^{\rm r}) = 5.5$. 
Reflecting the $\pi$-phase difference of the reflected microwave field, 
the two curves are out of phase, and their difference corresponds 
to the fidelity of the qubit readout, 
which is maximized at $-2.0$ rad and 1.2 rad to be 89\%. 

We further maximized the fidelity by tuning $t_{\rm p}$ and 
adding a short spike in the beginning of the readout pulse, 
and measured Rabi oscillations as shown in Fig.~\ref{fig4}d, 
where the length $t_{\rm c}$ and the amplitude $\sqrt{P_{\rm c}}$ 
of the qubit control pulse are swept. 
Figure~\ref{fig4}e shows the Rabi-oscillation measurement along the 
dashed line in Fig.~\ref{fig4}d. 
The contrast of the Rabi oscillations is 90.7\%. 
We attribute the sources of the error 
to incomplete initialization of the qubit by 2.6\% and 
to qubit energy relaxation (including the gate error of the $\pi$ pulse), which adds 6.7\%. 
Note that the non-zero minimum of the blue curve in Fig.~\ref{fig4}c is 
due to incomplete initialization of the qubit and not the non-locking error. 
The non-locking error is confirmed to be negligible for both states of the qubit 
(see details for Supplementary Information). 

As stated above, $N^{\rm r}$ of 5.5 is large enough to 
make the non-locking error of PPLO negligible. 
It is at the same time small enough for the qubit to avoid readout backaction. 
By sweeping $t_{\rm p}$, we can measure $T_1$ of the qubit, 
while populating photons in the readout resonator. 
It is measured to be 690~ns (see Supplementary Information), 
which agrees with $T_1$ obtained from the independent 
ensemble-averaged measurement using a standard pulse sequence, 
namely, a $\pi$ pulse followed by delayed readout. 
This indicates the nondestructive character of the present readout scheme. 

Another characteristic of the present readout scheme is its latching property. 
Once the qubit state is mapped to the oscillator state of PPLO, 
the information can be maintained, regardless of a subsequent 
qubit state transition, as long as the pump is turned on. 
We demonstrate this in Fig.~\ref{fig4}f, 
in which the readout fidelity is plotted as a function of $t_{\rm d}$. 
Even at $t_{\rm d}=6~\mu$s, at which the qubit has totally decayed, 
we do not lose readout fidelity. 

The present scheme enables fast, latching-type 
and single-shot readout of the qubit. 
In this sense, it is similar to schemes using a 
Josephson bifurcation amplifier,~\cite{Siddiqi06} 
Josephson chirped amplifier,~\cite{Naaman08} and period-doubling bifurcation,~\cite{Zorin11}
where the qubit is directly coupled to a nonlinear resonator. 
However, in the present scheme, 
the mean photon number in the resonator to which the qubit is coupled 
can be kept small (of order unity) regardless of the result of the readout. 
In this sense, it is similar to schemes where a qubit is coupled to a linear resonator, 
followed by an ultra low-noise amplifier such as JPA 
to achieve single-shot readout.~\cite{Lin13} 
Thus, the present scheme has combined advantages of 
both linear and nonlinear resonators, 
and can be useful in quantum error-correction protocols such as the surface code.~\cite{Fowler12} 

\vspace{5mm}
\small
\noindent\textbf{Acknowledgements} The authors would like to thank Y. Yamamoto, 
S. Utsunomiya, T. Kato, D. Vion and I. Mahboob for fruitful discussions. 
The authors are grateful to V. Bolkhovsky and G. Fitch for assistance 
with the device fabrication at MIT-LL. 
This work was partly supported by the Funding Program for World-Leading Innovative R\&D 
on Science and Technology (FIRST), Project for Developing Innovation Systems of MEXT, 
MEXT KAKENHI (Grant Nos. 21102002, 25400417), SCOPE (111507004), 
and National Institute of Information and Communications Technology (NICT).
\vspace{5mm}
\\
\noindent\textbf{Author Contributions} 
T.Y. and Z.R.L. conceived the experiment and wrote the manuscript. 
Z.R.L. performed the measurement and analyzed the data. 
K.I. fabricated the qubit device and helped the measurement and the analysis. 
T.Y. designed PPLO, which was fabricated at the group of W.D.O. 
K.K. and T.Y. developed the theory and performed numerical simulations. 
All authors contributed to the discussion of the results and helped editing the manuscript.
\vspace{5mm}
\\
\textbf{Additional Information}
The authors declare that they have no competing financial interests.
Supplementary information is available in the online version of the paper.


\clearpage

\begin{thebibliography}{10}
\expandafter\ifx\csname url\endcsname\relax
  \def\url#1{\texttt{#1}}\fi
\expandafter\ifx\csname urlprefix\endcsname\relax\def\urlprefix{URL }\fi
\providecommand{\bibinfo}[2]{#2}
\providecommand{\eprint}[2][]{\url{#2}}

\bibitem{Onyshkevych59}
\bibinfo{author}{Onyshkevych, L.~S.}, \bibinfo{author}{Kosonocky, W.~F.} \&
  \bibinfo{author}{Lo, A.~W.}
\newblock \bibinfo{title}{Parametric phase-locked oscillator ---
  characteristics and applications to digital systems}.
\newblock \emph{\bibinfo{journal}{Trans. Inst. Radio Engrs.}}
  \textbf{\bibinfo{volume}{EC-8}}, \bibinfo{pages}{277-286} (\bibinfo{year}{1959}).

\bibitem{Goto59}
\bibinfo{author}{Goto, E.}
\newblock \bibinfo{title}{The parametron, a digital computing element which
  utilizes parametric oscillation}.
\newblock \emph{\bibinfo{journal}{Proc. Inst. Radio Engrs.}}
  \textbf{\bibinfo{volume}{47}}, \bibinfo{pages}{1304-1316} (\bibinfo{year}{1959}).

\bibitem{Mahboob08}
\bibinfo{author}{Mahboob, I.} \& \bibinfo{author}{Yamaguchi, H.}
\newblock \bibinfo{title}{Bit storage and bit flip operations in an
  elecromechanical oscillator}.
\newblock \emph{\bibinfo{journal}{Nat. Nanotechnol.}}
  \textbf{\bibinfo{volume}{3}}, \bibinfo{pages}{275} (\bibinfo{year}{2008}).

\bibitem{Wang13}
\bibinfo{author}{Wang, Z.}, \bibinfo{author}{Marandi, A.},
  \bibinfo{author}{Wen, K.}, \bibinfo{author}{Byer, R.~L.} \&
  \bibinfo{author}{Yamamoto, Y.}
\newblock \bibinfo{title}{Coherent Ising machine based on degenerate optical
  parametric oscillators}.
\newblock \emph{\bibinfo{journal}{Phys. Rev. A}} \textbf{\bibinfo{volume}{88}},
  \bibinfo{pages}{063853} (\bibinfo{year}{2013}).

\bibitem{Siddiqi06}
\bibinfo{author}{Siddiqi, I.} \emph{et~al.}
\newblock \bibinfo{title}{Dispersive measurements of superconducting qubit
  coherence with a fast latching readout}.
\newblock \emph{\bibinfo{journal}{Phys. Rev. B}} \textbf{\bibinfo{volume}{73}},
  \bibinfo{pages}{054510} (\bibinfo{year}{2006}).

\bibitem{Vijay11}
\bibinfo{author}{Vijay, R.}, \bibinfo{author}{Slichter, D.~H.} \&
  \bibinfo{author}{Siddiqi, I.}
\newblock \bibinfo{title}{Observation of quantum jumps in a superconducting
  artificial atom}.
\newblock \emph{\bibinfo{journal}{Phys. Rev. Lett.}}
  \textbf{\bibinfo{volume}{106}}, \bibinfo{pages}{110502}
  (\bibinfo{year}{2011}).

\bibitem{Wallquist06}
\bibinfo{author}{Wallquist, M.}, \bibinfo{author}{Shumeiko, V.~S.} \&
  \bibinfo{author}{Wendin, G.}
\newblock \bibinfo{title}{Selective coupling of superconducting charge qubits
  mediated by a tunable stripline cavity}.
\newblock \emph{\bibinfo{journal}{Phys. Rev. B}} \textbf{\bibinfo{volume}{74}},
  \bibinfo{pages}{224506} (\bibinfo{year}{2006}).

\bibitem{Yamamoto08}
\bibinfo{author}{Yamamoto, T.} \emph{et~al.}
\newblock \bibinfo{title}{Flux-driven Josephson parametric amplifier}.
\newblock \emph{\bibinfo{journal}{Appl.\ Phys.\ Lett.}}
  \textbf{\bibinfo{volume}{93}}, \bibinfo{pages}{042510}
  (\bibinfo{year}{2008}).

\bibitem{Lin13}
\bibinfo{author}{Lin, Z.~R.} \emph{et~al.}
\newblock \bibinfo{title}{Single-shot readout of a superconducting flux qubit
  with a flux-driven Josephson parametric amplifier}.
\newblock \emph{\bibinfo{journal}{Appl.\ Phys.\ Lett.}}
  \textbf{\bibinfo{volume}{103}}, \bibinfo{pages}{132602}
  (\bibinfo{year}{2013}).

\bibitem{Wilson10}
\bibinfo{author}{Wilson, C.~M.} \emph{et~al.}
\newblock \bibinfo{title}{Photon generation in an electromagnetic cavity with a
  time-dependent boundary}.
\newblock \emph{\bibinfo{journal}{Phys. Rev. Lett.}}
  \textbf{\bibinfo{volume}{105}}, \bibinfo{pages}{233907}
  (\bibinfo{year}{2010}).

\bibitem{Krantz13}
\bibinfo{author}{Krantz, P.} \emph{et~al.}
\newblock \bibinfo{title}{Investigation of nonlinear effects in Josephson
  parametric oscillators used in circuit quantum electrodynamics}.
\newblock \emph{\bibinfo{journal}{New Journal of Physics}}
  \textbf{\bibinfo{volume}{15}}, \bibinfo{pages}{105002}
  (\bibinfo{year}{2013}).

\bibitem{Ryvkine06}
\bibinfo{author}{Ryvkine, D.} \& \bibinfo{author}{Dykman, M.~I.}
\newblock \bibinfo{title}{Resonant symmetry lifting in a parametrically
  modulated oscillator}.
\newblock \emph{\bibinfo{journal}{Phys. Rev. E}} \textbf{\bibinfo{volume}{74}},
  \bibinfo{pages}{061118} (\bibinfo{year}{2006}).

\bibitem{Kim10}
\bibinfo{author}{Kim, Y.} \emph{et~al.}
\newblock \bibinfo{title}{Observation of resonant symmetry lifting by an
  effective bias field in a parametrically modulated atomic trap}.
\newblock \emph{\bibinfo{journal}{Phys. Rev. A}} \textbf{\bibinfo{volume}{82}},
  \bibinfo{pages}{063407} (\bibinfo{year}{2010}).

\bibitem{Mahboob10}
\bibinfo{author}{Mahboob, I.}, \bibinfo{author}{Froitier, C.} \&
  \bibinfo{author}{Yamaguchi, H.}
\newblock \bibinfo{title}{A symmetry-breaking electromechanical detector}.
\newblock \emph{\bibinfo{journal}{Appl.\ Phys.\ Lett.}}
  \textbf{\bibinfo{volume}{96}}, \bibinfo{pages}{213103}
  (\bibinfo{year}{2010}).

\bibitem{Dykman98}
\bibinfo{author}{Dykman, M.~I.}, \bibinfo{author}{Maloney, C.~M.},
  \bibinfo{author}{Smelyanskiy, V.~N.} \& \bibinfo{author}{Silverstein, M.}
\newblock \bibinfo{title}{Fluctuational phase-flip transitions in
  parametrically driven oscillators}.
\newblock \emph{\bibinfo{journal}{Phys. Rev. E}} \textbf{\bibinfo{volume}{57}},
  \bibinfo{pages}{5202--5212} (\bibinfo{year}{1998}).

\bibitem{Johansson13}
\bibinfo{author}{Johansson, J.}, \bibinfo{author}{Nation, P.} \&
  \bibinfo{author}{Nori, F.}
\newblock \bibinfo{title}{Qutip 2: A Python framework for the dynamics of open
  quantum systems}.
\newblock \emph{\bibinfo{journal}{Computer Physics Communications}}
  \textbf{\bibinfo{volume}{184}}, \bibinfo{pages}{1234 -- 1240}
  (\bibinfo{year}{2013}).

\bibitem{AndersonBook}
\bibinfo{author}{Anderson, J.~B.}, \bibinfo{author}{Aulin, T.} \&
  \bibinfo{author}{Sundberg, C.-E.}
\newblock \emph{\bibinfo{title}{Digital Phase Modulation}}
  (\bibinfo{publisher}{Plenum Press}, \bibinfo{address}{New York},
  \bibinfo{year}{1986}).

\bibitem{Inomata12}
\bibinfo{author}{Inomata, K.}, \bibinfo{author}{Yamamoto, T.},
  \bibinfo{author}{Billangeon, P.-M.}, \bibinfo{author}{Nakamura, Y.} \&
  \bibinfo{author}{Tsai, J.~S.}
\newblock \bibinfo{title}{Large dispersive shift of cavity resonance induced by
  a superconducting flux qubit in the straddling regime}.
\newblock \emph{\bibinfo{journal}{Phys. Rev. B}} \textbf{\bibinfo{volume}{86}},
  \bibinfo{pages}{140508} (\bibinfo{year}{2012}).

\bibitem{Naaman08}
\bibinfo{author}{Naaman, O.}, \bibinfo{author}{Aumentado, J.},
  \bibinfo{author}{Friedland, L.}, \bibinfo{author}{Wurtele, J.~S.} \&
  \bibinfo{author}{Siddiqi, I.}
\newblock \bibinfo{title}{Phase-locking transition in a chirped superconducting
  Josephson resonator}.
\newblock \emph{\bibinfo{journal}{Phys. Rev. Lett.}}
  \textbf{\bibinfo{volume}{101}}, \bibinfo{pages}{117005}
  (\bibinfo{year}{2008}).

\bibitem{Zorin11}
\bibinfo{author}{Zorin, A.~B.} \& \bibinfo{author}{Makhlin, Y.}
\newblock \bibinfo{title}{Period-doubling bifurcation readout for a Josephson
  qubit}.
\newblock \emph{\bibinfo{journal}{Phys. Rev. B}} \textbf{\bibinfo{volume}{83}},
  \bibinfo{pages}{224506} (\bibinfo{year}{2011}).

\bibitem{Fowler12}
\bibinfo{author}{Fowler, A.~G.}, \bibinfo{author}{Mariantoni, M.},
  \bibinfo{author}{Martinis, J.~M.} \& \bibinfo{author}{Cleland, A.~N.}
\newblock \bibinfo{title}{Surface codes: Towards practical large-scale quantum
  computation}.
\newblock \emph{\bibinfo{journal}{Phys. Rev. A}} \textbf{\bibinfo{volume}{86}},
  \bibinfo{pages}{032324} (\bibinfo{year}{2012}).

\end{thebibliography}
\end{document}


{{
\title{Supplementary information: Josephson parametric phase-locked oscillator and its
application to dispersive readout of superconducting qubits}

\author{Z. R. Lin} 
\affiliation{RIKEN Center for Emergent Matter Science (CEMS), Wako, Saitama 351-0198, Japan}

\author{K. Inomata} 
\affiliation{RIKEN Center for Emergent Matter Science (CEMS), Wako, Saitama 351-0198, Japan}

\author{K. Koshino}
\affiliation{College of Liberal Arts and Sciences, Tokyo Medical and Dental University, 
Ichikawa, Chiba 272-0827, Japan}

\author{W. D. Oliver}
\affiliation{MIT Lincoln Laboratory, Lexington, Massachusetts 02420, USA}

\author{Y. Nakamura}
\affiliation{Research Center for Advanced Science and Technology (RCAST), 
The University of Tokyo, Meguro-ku, Tokyo 153-8904, Japan}
\affiliation{RIKEN Center for Emergent Matter Science (CEMS), Wako, Saitama 351-0198, Japan}
 
\author{J. S. Tsai}
\affiliation{RIKEN Center for Emergent Matter Science (CEMS), Wako, Saitama 351-0198, Japan}
\affiliation{NEC Smart Energy Research Laboratories, Tsukuba, Ibaraki 305-8501, Japan}

\author{T. Yamamoto} \email[t-yamamoto@fe.jp.nec.com]{}
\affiliation{RIKEN Center for Emergent Matter Science (CEMS), Wako, Saitama 351-0198, Japan}
\affiliation{NEC Smart Energy Research Laboratories, Tsukuba, Ibaraki 305-8501, Japan}


\maketitle

\nopagebreak

\section{Experimental details}
\subsection{Images of the Josephson parametric phase-locked oscillator}
Figure \ref{FigS_device}a shows the device image of the parametric 
phase-locked oscillator (PPLO).
The device consists of a quarter wavelength (cavity length 2.6~mm) coplanar waveguide 
(CPW) resonator with a dc-SQUID termination and a 
pump line inductively coupled to the SQUID loop (mutual inductance $M \sim 1.0$~pH).
The critical current of the Josephson junction of the SQUID 
is estimated to be 3.1~$\mu$A for each junction from 
the fitting shown in Fig.~\ref{FigS_fr_flux}. 
The device was fabricated by the planarized niobium trilayer 
process at MIT Lincoln laboratory.
The resonator and the pump line are made of 150-nm-thick niobium film 
sputtered on a Si substrate covered by 500-nm-thick SiO${}_2$ layer. 
Figure \ref{FigS_device}b shows the magnified image of 
the coupling capacitance between the microwave feedline and the resonator $C_{\rm in}^{\rm PO}$, 
which is designed to be 15~fF. 
The internal and external quality factors are measured to be 
5200 and 340, respectively at $\omega_0^{\rm PO}/2\pi=10.507$~GHz, 
which give the loaded quality factor $Q_{\rm L}^{\rm PO}$ of 320. 
Figures~\ref{FigS_device}c and d show 
the magnified optical image and the scanning electron micrograph 
of the dc-SQUID part, respectively. 

\subsection{Error budget of the Rabi-oscillation contrast}
In the main article, we show Rabi oscillations with a contrast of 90.7\%. 
Here, we present our analysis on the loss of the contrast.
Possible sources of the error are 
(i) incomplete initialization of the qubit, (ii) insufficient power of the locking signal (LS), 
and (iii) qubit energy relaxation (including the gate error in the qubit-control $\pi$ pulse). 

The error from the first source (incomplete initialization) 
is estimated from the direct measurement of the background qubit excitation 
by operating the device as Josephson parametric amplifier (JPA)~\cite{Lin13}, 
and found to be 2.6\%. 
Half of this is equal to the non-zero minimum of the blue curve in 
Fig.~4c in the main article. 
This indicates that the error from the second source (non-locking error) 
is negligible at least when the qubit is in state $|0\rangle$. 

Non-locking error is also estimated from the measurement result shown 
in Fig.~\ref{FigS_nonlock_err}, 
which is similar to the measurement shown in Fig.~2e in the main article, 
and found to be negligible for both of the qubit states. 
Minimum probability of $1\pi$ state decreases as we increase 
$P_{\rm s}^{\rm PO}$ and saturates at the level determined by the background qubit excitation 
when $P_{\rm s}^{\rm PO}$ is larger than $\sim -124$~dBm. 

The microwave powers injected into the PPLO when the qubit is in $|0\rangle$ state 
and in $|1\rangle$ state are shown 
in the figure as $P_{\rm s}^0$ and $P_{\rm s}^1$, respectively. 
Because of the dispersive shift in $\omega_0^{\rm r}$ and 
finite internal loss of the readout resonator, 
$P_{\rm s}^1$ is slightly lower than $P_{\rm s}^0$ which is set at $-122.5$~dBm. 
$P_{\rm s}^1$ is estimated by directly measuring the amplitude of 
the reflected microwave when the qubit is excited to $|1\rangle$ state. 
Figure \ref{FigS_P0P1} shows the histograms of the voltage of the 
reflected readout pulse when the qubit-control $\pi$ pulse is not 
applied (a) and is applied (b) to the qubit. 
The reflected voltage is measured by a single-shot readout using JPA~\cite{Lin13}. 
The peak in Fig.~\ref{FigS_P0P1}a corresponds to the qubit state $|0\rangle$,
while the right peak in Fig.~\ref{FigS_P0P1}b 
corresponds to the qubit state $|1\rangle$. 
By fitting the peaks with Gaussian functions, we extract the voltage of the 
reflected readout pulse when the qubit is in $|0\rangle$ state and 
in $|1\rangle$ state, which we call $V_0$ and $V_1$, respectively. 
$V_i$ and $P_{\rm s}^i$ ($i$=0,1) satisfy the following relation 
\begin{equation}~\label{V_P}
\frac{P_{\rm s}^{0}G(P_{\rm s}^{0})}{P_{\rm s}^{1}G(P_{\rm s}^{1})} = 
\Bigl( \frac{V_0}{V_1} \Bigr)^2,
\end{equation}
where $G$ represents the power gain of the JPA. 
Note that $G$ is not necessarily independent of the input power 
when the input power is high, 
and we measured it independently (data not shown). 
From Eq.~\ref{V_P}, $P_{\rm s}^{1}$ is estimated to be $-123.0$~dBm. 

The rest of the error is attributable to the third source (qubit relaxation). 
The energy relaxation time $T_1$ of the qubit is measured to be 690~ns (see below). 
Assuming $1-\exp(-t_{\rm w}/T_1)=0.067$, 
$t_{\rm w}=48$~ns is obtained, which is close to the sum of 
$t_{\rm c}$, $t_{\rm p}$ and the response time of PPLO measured 
to be $\sim10$~ns (Fig.~\ref{FigS_tcon}).

\subsection{Qubit energy relaxation time}
Figure \ref{FigS_T1} shows the probability of 0$\pi$ state as a function of 
$t_{\rm p}$ when the qubit-control $\pi$ pulse is turned off and on. 
We fit the data ($\pi$ pulse on) 
from $t_{\rm p}=40$ ns to 3 $\mu$s with an exponential function, 
and obtain the time constant of 690 ns. 
This agrees with the qubit energy relaxation time $T_1$ of $694$ ns
obtained from an independent ensemble-averaged measurement 
using standard pulse sequence for $T_1$ measurement, namely, 
$\pi$ pulse followed by the delayed readout. 
Small decay of the data without $\pi$ pulse is possibly due to the 
qubit excitation induced by the readout pulse. 

\section{Theory and simulations}
\subsection{Hamiltonian and equations of motion}
The Hamiltonian of the PPLO including a signal port for the locking signal (LS) and 
a fictitious loss port for internal loss of the resonator is given by~\cite{Wallsbook}
\begin{eqnarray} \label{hamiltonian}
\mathcal{H}(t) &=& \mathcal{H}_{\rm sys}(t) + \mathcal{H}_{\rm sig} + \mathcal{H}_{\rm loss}, \\
\mathcal{H}_{\rm sys}(t)/\hbar &=& \omega_0^{\rm PO} \bigl[ a^\dagger a + 
\epsilon \cos(\omega_{\rm p} t) (a + a^\dagger)^2 \bigr] + \gamma (a + a^\dagger)^4, \\
\mathcal{H}_{\rm sig}/\hbar &=& \int dk \Bigl[v_b k b_k^\dagger b_k + 
\mathrm{i} \sqrt{\frac{v_b \kappa_1}{2 \pi}}
\Bigl( a^\dagger b_k-b_k^\dagger a \Bigr) \Bigr], \\
\mathcal{H}_{\rm loss}/\hbar &=& \int dk \Bigl[v_c k c_k^\dagger c_k + 
\mathrm{i} \sqrt{\frac{v_c \kappa_2}{2 \pi}}
\Bigl( a^\dagger c_k-c_k^\dagger a \Bigr) \Bigr], 
\end{eqnarray}
where $\omega_0^{\rm PO}$ is the static resonant frequency of the PPLO, 
$\omega_{\rm p}$ and $\epsilon$ represent the frequency and 
magnitude of the parametric modulation, respectively, 
$\gamma$ represents the nonlinearity of the Josephson junction (JJ), 
$a$ is the annihilation operator for the resonator, 
$b_k$ ($c_k$) is the annihilation operator for the photon in the signal (loss) port 
with a wave number $k$ and a velocity $v_b$ ($v_c$), and 
$\kappa_1$ ($\kappa_2$) represents the coupling strength between the resonator 
and the signal (loss) port. 
The operators satisfy the following commutation rules: 
$\bigl[ a,a^\dagger \bigr] = 1$, $\bigl[ b_k,{b_k'}^\dagger \bigr] = \delta (k-k')$, and 
$\bigl[ c_k,{c_k'}^\dagger \bigr] = \delta(k-k')$. 
The coupling constants $\kappa_1$ and $\kappa_2$ are related to the 
external and internal quality factors of the resonator as follows: 
$\kappa_1 = \omega_0^{\rm PO}/Q_{\rm e}^{\rm PO}$ and 
$\kappa_2 = \omega_0^{\rm PO}/Q_{\rm i}^{\rm PO}$. 
Below we consider the case where $\omega_{\rm p} = 2\omega_0^{\rm PO}$. 

From the Heisenberg equations of motion for $b_k$, we obtain 
\begin{equation}\label{Heq_b}
\frac{db_k(t)}{dt} = -\mathrm{i} vk b_k(t) - \sqrt{\frac{v\kappa_1}{2 \pi}} a(t). 
\end{equation} 
By solving this differential equation formally, 
we have
\begin{equation}\label{sol_Heq_b}
b_k(t) = e^{- \mathrm{i} vkt} b_k(0) 
- \sqrt{\frac{v\kappa_1}{2 \pi}} \int_{0}^{t} e^{- \mathrm{i} vk(t - t')} a(t')dt'. 
\end{equation} 
We introduce the real-space representation of the waveguide field by
$\widetilde{b}_r=(2\pi)^{-1/2}\int dk e^{\mathrm{i}kr} b_k$.
In this representation, 
the waveguide field interacts with the resonator at $r=0$ and 
the $r<0$ ($r>0$) region corresponds to the incoming (outgoing) field.
From Eq.~\ref{sol_Heq_b}, we have
\begin{eqnarray}
\widetilde{b}_r(t) 
&=& \widetilde{b}_{r-vt}(0) 
- \sqrt{\frac{\kappa_1}{v}} \theta(r)\theta(t-r/v)a(t-r/v) \label{br_op}, 
\end{eqnarray}
where $\theta(r)$ is the Heviside step function. 
We define the input and output operators by 
\begin{eqnarray} \label{bin_op}
\widetilde{b}_{\rm in}(t) &\equiv& \widetilde{b}_{-0}(t) 
= \widetilde{b}_{-vt}(0), \\
\widetilde{b}_{\rm out}(t) &\equiv& \widetilde{b}_{+0}(t) 
= \widetilde{b}_{\rm in}(t) - \sqrt{\frac{\kappa_1}{v}} a(t). 
\end{eqnarray}
Using Eqs.~\ref{br_op} and \ref{bin_op}, 
the field operator $\widetilde{b}_r(t)$ at the resonator position ($r=0$) is given by 
\begin{equation} \label{b0_op}
\widetilde{b}_0(t) = \frac{1}{\sqrt{2\pi}}\int b_k(t)dk = 
\widetilde{b}_{\rm in}(t) - \frac{1}{2}\sqrt{\frac{\kappa_1}{v}} a(t). 
\end{equation}

From the Heisenberg equations of motion for $a$, we obtain 
\begin{equation}\label{Heq_a}
\frac{da}{dt} = 
-\mathrm{i}[\mathcal{H}_{\rm sys}(t),a]
+ \sqrt{v\kappa_1} \widetilde{b}_0
+ \sqrt{v\kappa_2} \widetilde{c}_0. 
\end{equation}
Using Eq.~\ref{b0_op} and its counterpart for $\widetilde{c}_0$, 
Eq.~\ref{Heq_a} is rewritten as 
\begin{equation}\label{Heq_a3}
\frac{da}{dt} = 
-\mathrm{i}[\mathcal{H}_{\rm sys}(t),a]-\frac{\kappa}{2}a
+ \sqrt{v\kappa_1} \widetilde{b}_{\rm in}(t) + 
\sqrt{v\kappa_2} \widetilde{c}_{\rm in}(t), 
\end{equation}
where $\kappa=\kappa_1 + \kappa_2$. 
Now we switch to a frame rotating at $\omega_0^{\rm PO}$
[namely,
$a(t)e^{\mathrm{i}\omega_0^{\rm PO} t} \to a(t)$,
$b_{\rm in}(t)e^{\mathrm{i}\omega_0^{\rm PO} t} \to b_{\rm in}(t)$, and
$c_{\rm in}(t)e^{\mathrm{i}\omega_0^{\rm PO} t} \to c_{\rm in}(t)$]
and drop the rapidly rotating terms in $\mathcal{H}_{\rm sys}(t)$.
Then the static system Hamiltonian is given by
\begin{equation} \label{hamiltonian_simple}
\mathcal{H}_{\rm sys}/\hbar = 
\frac{\epsilon \omega_0^{\rm PO}}{2}(a^2 + a^{\dagger 2}) 
+ 6\gamma a^\dagger a^\dagger a a. 
\end{equation}
Here, we neglected the term $12\gamma a^\dagger a$
which can be regraded as a small renormalization to $\omega_0^{\rm PO}$.
From Eqs.~\ref{Heq_a3} and \ref{hamiltonian_simple}, we have
\begin{equation}\label{Heq_a4}
\frac{da}{dt} = 
-\frac{\kappa}{2}a 
-\mathrm{i} \omega_0^{\rm PO} \epsilon a^\dagger 
-12\mathrm{i}\gamma a^\dagger aa
+ \sqrt{v\kappa_1} \widetilde{b}_{\rm in}(t) + 
\sqrt{v\kappa_2} \widetilde{c}_{\rm in}(t).
\end{equation}

Now we consider the classical amplitude of the resonator field, namely, 
$\langle a(t) \rangle$.
We denote the locking signal applied from the signal port by
$E_{\rm s}(r,t)=E_{\rm s}^* e^{\mathrm{i}\omega_0^{\rm PO}(r/v-t)}$. 
Then, we rigorously have
$\langle \widetilde{b}_{\rm in}(t) \rangle = E_{\rm s}^*$ 
and $\langle \widetilde{c}_{\rm in}(t) \rangle = 0$.
Dividing $\langle a(t)\rangle$ into its quadratures as
$\langle a(t)\rangle = q_x(t) - \mathrm{i}q_y(t)$,
their equations of motion are given by
\begin{eqnarray}
\frac{dq_x}{dt} &=& -\frac{\kappa}{2}q_x + \omega_0^{\rm PO} \epsilon q_y 
- 12\gamma(q_x^2+q_y^2)q_y + \sqrt{\kappa_1}
|E_{\rm s}|\cos\theta_{\rm s},
\label{eq_of_mot_7} \\
\frac{dq_y}{dt} &=& -\frac{\kappa}{2}q_y + \omega_0^{\rm PO} \epsilon q_x 
+ 12\gamma(q_x^2+q_y^2)q_x + \sqrt{\kappa_1}
|E_{\rm s}|\sin\theta_{\rm s},
\label{eq_of_mot_8}
\end{eqnarray}
where $E_{\rm s} = |E_{\rm s}|e^{\mathrm{i}\theta_{\rm s}}$, 
and we approximated $\langle a^\dagger aa \rangle$ to be 
$\langle a^\dagger \rangle \langle a \rangle^2$. 
Equations~\ref{eq_of_mot_7} and \ref{eq_of_mot_8} can be recast as 
\begin{eqnarray}
\frac{dq_x}{dt} &=& -\frac{\kappa}{2}q_x + \frac{\partial g}{\partial q_y}, \\
\frac{dq_y}{dt} &=& -\frac{\kappa}{2}q_y - \frac{\partial g}{\partial q_x}, 
\end{eqnarray}
where 
\begin{equation}
g(q_x,q_y) = \frac{\epsilon}{2}\omega_0^{\rm PO}(q_y^2 - q_x^2) - 3\gamma(q_x^2 + q_y^2)^2 + 
\sqrt{\kappa_1} |E_{\rm s}| (q_y \cos\theta_{\rm s} - q_x \sin\theta_{\rm s}). 
\end{equation}
Considering that $P_{\rm p}/P_{\rm p0}=(\epsilon/\epsilon_0)^2$, 
where $\epsilon_0=\kappa/(2\omega_0^{\rm PO})$ is the threshold for $\epsilon$, 
these are Eqs.~2 to 4 in the main article. 

\subsection{Master equation}
We denote the resonator transition operator by $s_{mn}=|m\rangle \langle n|$,
where $|m\rangle$ and $|n\rangle$ are the Fock states.
Its Heisenberg equation is given,
in the rotating frame [$s_{mn}(t)e^{\mathrm{i}\omega_0^{\rm PO}(n-m)t} \to s_{mn}(t)$], by
\begin{eqnarray}
\frac{d}{dt}s_{mn} &=& 
\frac{\mathrm{i}}{\hbar}[\mathcal{H}_{\rm sys},s_{mn}] 
+ \frac{\kappa}{2}(2a^\dagger s_{mn}a - s_{mn}a^\dagger a - a^\dagger as_{mn}) \nonumber \\
&& + \sqrt{v\kappa_1}[s_{mn},a^\dagger]\widetilde{b}_{\rm in}(t) 
- \sqrt{v\kappa_1}\widetilde{b}_{\rm in}^\dagger(t)[s_{mn},a] \nonumber \\
&& + \sqrt{v\kappa_2}[s_{mn},a^\dagger]\widetilde{c}_{\rm in}(t) 
- \sqrt{v\kappa_2}\widetilde{c}_{\rm in}^\dagger(t)[s_{mn},a]. 
\end{eqnarray}
where the static system Hamiltonian $\mathcal{H}_{\rm sys}$ 
is given by Eq.~\ref{hamiltonian_simple}.
Using again that 
$\langle \widetilde{b}_{\rm in}(t) \rangle = E_{\rm s}^*$ 
and $\langle \widetilde{c}_{\rm in}(t) \rangle = 0$,
$\langle s_{mn} \rangle$ evolves as
\begin{eqnarray}~\label{Heq_Bexp}
\frac{d}{dt}\langle s_{mn} \rangle &=& 
\frac{\mathrm{i}\omega_0^{\rm PO}\epsilon}{2} \Bigl( 
\sqrt{m(m-1)}\langle s_{m-2,n} \rangle + \sqrt{(m+1)(m+2)} \langle s_{m+2,n} \rangle \nonumber \\
&& - \sqrt{n(n-1)}\langle s_{m,n-2} \rangle - \sqrt{(n+1)(n+2)} \langle s_{m,n+2} \rangle
\Bigr) \nonumber \\
&& + 6\mathrm{i}\gamma[m(m-1)-n(n-1)] \langle s_{mn} \rangle 
+ \frac{\kappa}{2} \Bigl[
2\sqrt{(m+1)(n+1)}\langle s_{m+1,n+1} \rangle - (m+n) \langle s_{mn} \rangle \Bigr] \nonumber \\
&& + \sqrt{\kappa_1} E_{\rm s} \Bigl(
\sqrt{n}\langle s_{m,n-1} \rangle - \sqrt{m+1}\langle s_{m+1,n} \rangle 
\Bigr) \nonumber \\
&& - \sqrt{\kappa_1} E_{\rm s}^* \Bigl( 
\sqrt{n+1}\langle s_{m,n+1} \rangle - \sqrt{m}\langle s_{m-1,n} \rangle
\Bigr),
\end{eqnarray}
where we have used $a^\dagger s_{mn}a = \sqrt{(m+1)(n+1)}s_{m+1,n+1}$ and similar equalities. 

Since $\langle s_{mn} \rangle = {\rm Tr}[\rho s_{mn}] = \rho_{nm}$, 
Eq.~\ref{Heq_Bexp} is equivalent to the following master equation, 
\begin{equation}~\label{dim_master}
\frac{d\rho}{dt} = -\frac{\mathrm{i}}{\hbar}[\mathcal{H}_{\rm int}, \rho] 
+ \frac{\kappa}{2}(2a\rho a^\dagger - a^\dagger a \rho - \rho a^\dagger a), 
\end{equation}
where 
\begin{equation}
\mathcal{H}_{\rm int}/\hbar = 
\frac{\epsilon \omega_0^{\rm PO}}{2}(a^2 + a^{\dagger 2})
+ 6\gamma a^\dagger a^\dagger aa 
+ \mathrm{i}\sqrt{\kappa_1}|E_{\rm s}|
(e^{\mathrm{i}\theta_{\rm s}}a^\dagger - e^{\mathrm{-i}\theta_{\rm s}}a). 
\end{equation}
By introducing dimensionless time $\tau = t\kappa/2$, 
Eq.~\ref{dim_master} becomes 
\begin{equation}~\label{master_eq}
\frac{d\rho}{d\tau} = -\mathrm{i}[\mathcal{H}'_{\rm int},\rho] + 
(2a\rho a^\dagger - a^\dagger a\rho - \rho a^\dagger a). 
\end{equation}
Here the dimensionless Hamiltonian $\mathcal{H}'_{\rm int}$ is given by 
\begin{equation}
\mathcal{H}'_{\rm int} = \mathcal{H}_{\rm int}/(\hbar\kappa/2)
= \gamma'a^\dagger a^\dagger aa + 
\frac{1}{2}\sqrt{\frac{P_{\rm p}}{P_{\rm p0}}}(a^2 + a^{\dagger2}) + 
\mathrm{i}\sqrt{N^{\rm PO}}
(e^{\mathrm{i}\theta_{\rm s}}a^\dagger - e^{\mathrm{-i}\theta_{\rm s}}a), 
\end{equation}
where $\gamma'=12\gamma/\kappa$, 
$\sqrt{N^{\rm PO}}=\sqrt{\kappa_1}|E_{\rm s}|/(\kappa/2)$. 

We numerically solve Eq.~\ref{master_eq} by 
expanding $\rho$ in the number state basis~\cite{Johansson13}, 
namely $\rho=\sum_{m,n=0}^N \rho_{mn}|m\rangle \langle n|$, 
and calculate the $Q$ function $Q(z)=\langle z|\rho|z \rangle/\pi$, 
where $z$ is a coherent state~\cite{Wallsbook}. 

\subsection{Experimental parameters}
To simulate the experiments, we need to determine the parameters 
such as $N^{\rm PO}$, $P_{\rm p0}$, and $\gamma$. 
$N^{\rm PO}$ is determined from the above definition and 
$|E_{\rm s}|=\sqrt{P_{\rm s}^{\rm PO}/\hbar \omega_0^{\rm PO}}$. 
Since we do not precisely know the mutual inductance between 
the pump line and the SQUID loop in the PPLO, we can only roughly 
determine $P_{\rm p0}$, the threshold for the pump power, 
from the measurement shown in Fig.~2a in the main article, 
and leave it as a semi-adjustable parameter. 
$\gamma$ is calculated based on the theory in Ref.~\onlinecite{Wallquist06}. 
It is related with the wavenumber $k$ of the first mode of the CPW resonator 
terminated by a SQUID, 
\begin{equation}~\label{gamma_Bk}
\gamma = -\Bigl( \frac{2\pi}{\Phi_0} \Bigr)^2\frac{\hbar B_k}{8C_k}, 
\end{equation} 
where 
\begin{eqnarray}~\label{B_k_param}
B_k &=& \frac{(1/4)\cos^2(kd)}{1+2kd/\sin(2kd)}, \\ \label{equivC}
C_k &=& \frac{C_{\rm cav}}{2}\Bigl[ 1+\frac{\sin(2kd)}{2kd} \Bigr] + C_{\rm J}\cos^2(kd). 
\end{eqnarray}
Here, $d$ and $C_{\rm cav}$ are the length and the total capacitance of the CPW resonator, 
respectively, and $C_{\rm J}$ is the total junction capacitance of the SQUID. 
The flux dependent resonant frequency is given by 
\begin{equation}~\label{fdep_omega0}
\omega_0^{\rm PO} (\Phi_{\rm sq}) 
= \frac{1}{\sqrt{L_k(C_k+C_{\rm in}^{\rm PO})}}, 
\end{equation}
where 
\begin{equation}
1/L_k = \frac{(kd)^2}{2L_{\rm cav}}\Bigl[ 1+\frac{\sin(2kd)}{2kd} + \frac{2C_{\rm J}}{C_{\rm cav}}
\cos^2(kd) \Bigr]. 
\end{equation}
Here, $L_{\rm cav}$ is the total inductance of the CPW resonator, and $kd$ is determined by 
the equation~\cite{Wallquist06}, 
\begin{equation}~\label{kd_eq}
kd\tan(kd) = \Bigl( \frac{2\pi}{\Phi_0} \Bigr) L_{\rm cav}2I_0\Bigl| 
\cos\Bigl( \pi \frac{\Phi_{\rm sq}}{\Phi_0} \Bigr) \Bigr| - 
\frac{C_{\rm J}}{C_{\rm cav}}(kd)^2, 
\end{equation}
where $I_0$ represents the critical current of each of the SQUID JJ. 

We fit the data shown in Fig.~\ref{FigS_fr_flux} by 
Eq.~\ref{fdep_omega0} with fitting parameters of $I_{\rm c}$ and $C_{\rm cav}$, 
which are determined to be 3.1~$\mu$A and 410~fF, respectively. 
Using Eqs.~\ref{gamma_Bk} to \ref{equivC}, and \ref{kd_eq}, 
we calculate $\gamma$ to be 
$-2.3\times10^5$ Hz at $\omega_0^{\rm PO}/2\pi=10.51$ GHz. 

\subsection{Simulation of non-locking error}
Figure~\ref{fig_th}a shows an example of the calculated $Q$ function. 
The parameters used in the calculation are 
$P_{\rm p}/P_{\rm p0}=1.660$ (2.2~dB), $N^{\rm PO}=0.09$, $\theta_{\rm s}=\pi/2$. 
The density matrix is truncated at $N=80$, which 
we confirmed large enough for these parameters. 
Also, the calculation is stopped at $\tau=20$, 
which we confirmed long enough for the system to become stationary. 
We clearly observe two distribution peaks, which correspond to 
$0\pi$ and $1\pi$ states of PPLO. 
Based on this result, we get the probability of $0\pi$ state 
by integrating $Q$ for $q_x > 0$. 
Figure~\ref{fig_th}b shows the probability of $0\pi$ state as a function of 
the LS phase for different $N^{\rm PO}$'s. 
$N^{\rm PO}$'s are chosen from those in Fig.~2d in the main article. 
The agreement is fairly good.

\newpage

\begin{figure}
\begin{center}
\includegraphics[width=0.75\columnwidth,clip]{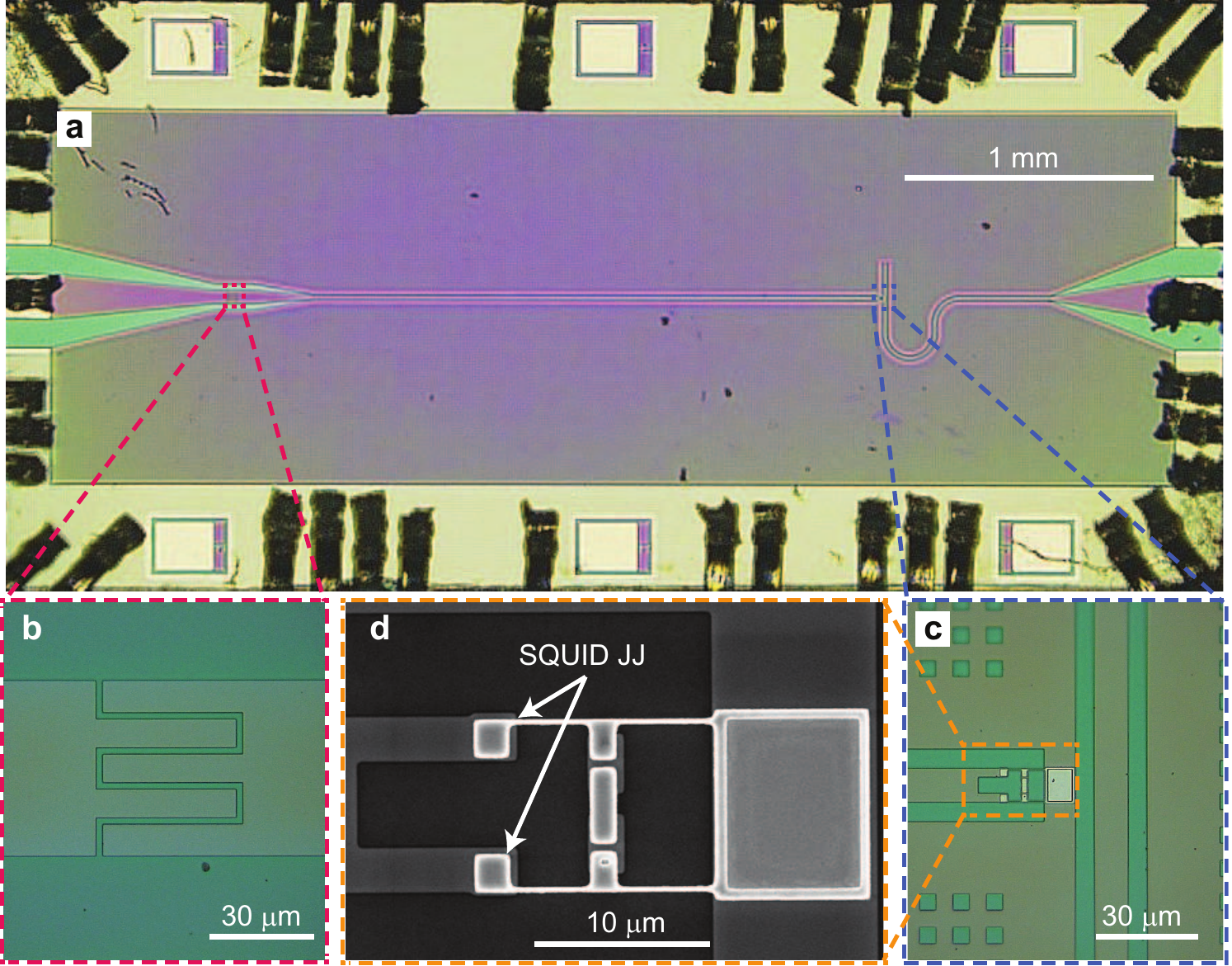}
\end{center}
\caption{Images of the parametric phase-locked oscillator. 
{\bf a,} Optical image of the device. 
The $\lambda/4$-type CPW resonator is made of 150-nm-thick niobium film
deposited on an oxidized silicon substrate. 
{\bf b,} Magnified image of the coupling capacitance. 
{\bf c,} Magnified optical image of the dc-SQUID part and 
{\bf d,} its scanning electron micrograph. 
The SQUID loop contains a three-Josephson-junction flux qubit, 
which is not used in this work.}
\label{FigS_device}
\end{figure}

\begin{figure}
\begin{center}
\includegraphics[width=0.56\columnwidth,clip]{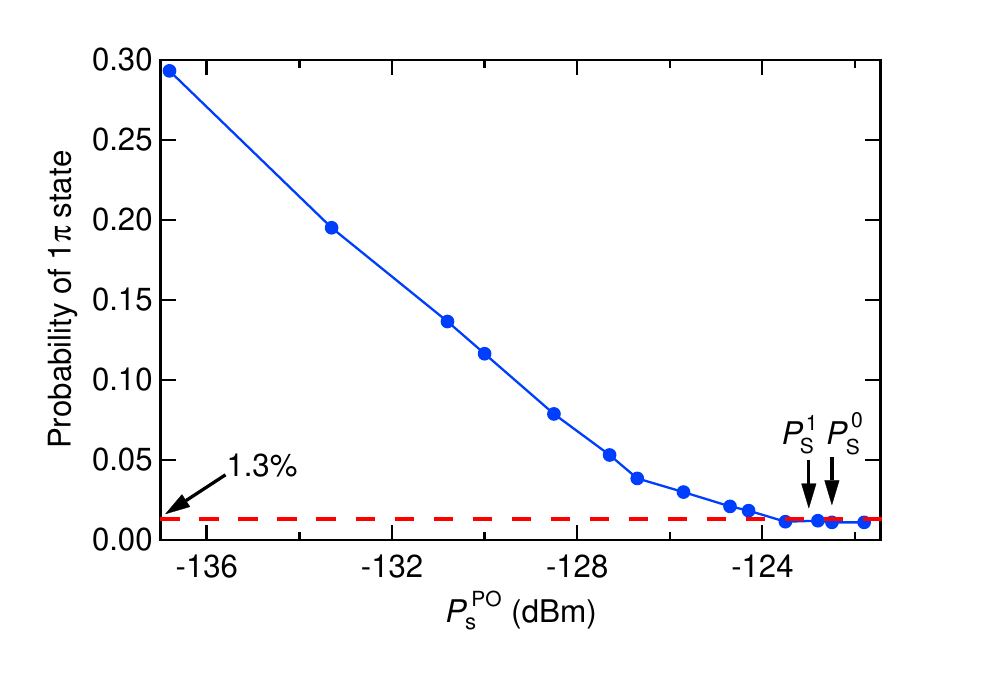}
\end{center}
\caption{Minimum probability of $1\pi$ state as a function of the locking-signal power. 
The non-zero minimum probability indicated by the horizontal dashed line 
is due to the initialization error of the qubit.
$P_{\rm s}^0$ and $P_{\rm s}^1$ represent the microwave powers 
injected into the PPLO when the qubit is in $|0\rangle$ state 
and in $|1\rangle$ state, respectively. 
}
\label{FigS_nonlock_err}
\end{figure}

\begin{figure}
\begin{center}
\includegraphics[width=0.56\columnwidth,clip]{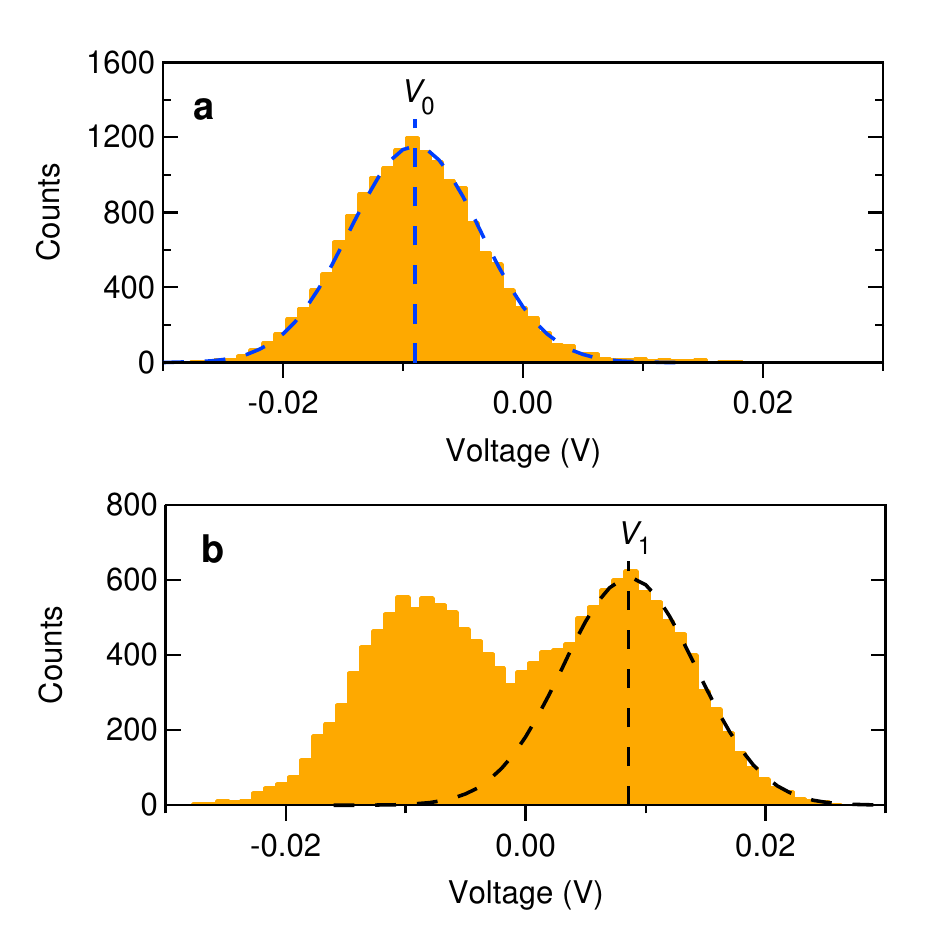}
\end{center}
\caption{Histograms of the voltage of the reflected readout pulse 
{\bf a}, with qubit-control $\pi$ pulse off and {\bf b}, on. 
The input power to the JPA is $-122.5$~dBm, 
which is the same as in the Rabi-oscillation measurement in the main article. 
The JPA is operated at a gain of 13 dB and a bandwidth of 21 MHz 
with 1-dB-compression point of $-123$~dBm.
The dashed lines represent the Gaussian fits to the distribution peaks 
to extract the mean values of $V_0=-0.090$~V and $V_1=0.086$~V. 
Only right-hand side of the peak is fitted for {\bf b}. 
}
\label{FigS_P0P1}
\end{figure}

\begin{figure}
\begin{center}
\includegraphics[width=0.56\columnwidth,clip]{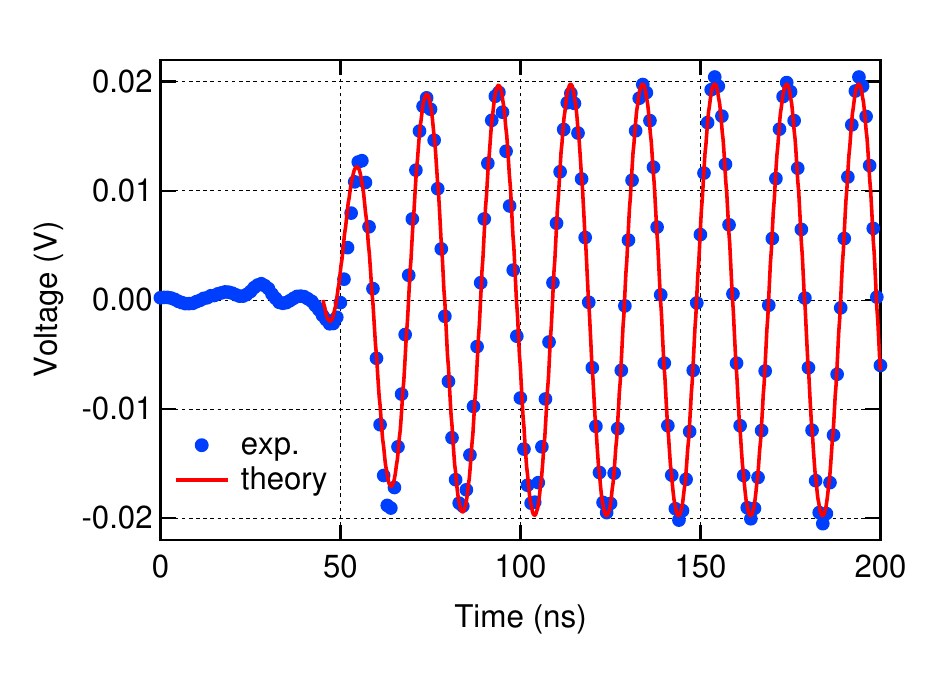}
\end{center}
\caption{Time trace of the PPLO output signal. 
$\omega_{\rm p}/2\pi=2\times10.193$~GHz, $t_{\rm p}=30$~ns, $t_{\rm r}=50$~ns, and 
$N^{\rm r}=5.5$ (See main article for definitions). 
The output signal at $\omega_{\rm p}/2$ is down-converted to 
$\omega_{\rm IF}=2\pi \times50$~MHz, 
and digitized at 1~GS/s. Qubit-control $\pi$ pulse is turned on ($t_{\rm c}=10$~ns). 
We selected 9250 traces which are in $1\pi$-phase state and 
averaged them (blue circles). 
Solid curve represents a fit to the data for $t>t_0=45$~ns. 
We assumed a function $V(t)=A(1-e^{-(t-t_0)/\tau})\cos(\omega_{\rm IF}t+\phi_0)$, 
and obtained the time constant $\tau$ of 9.4~ns. 
}
\label{FigS_tcon}
\end{figure}

\begin{figure}
\begin{center}
\includegraphics[width=0.56\columnwidth,clip]{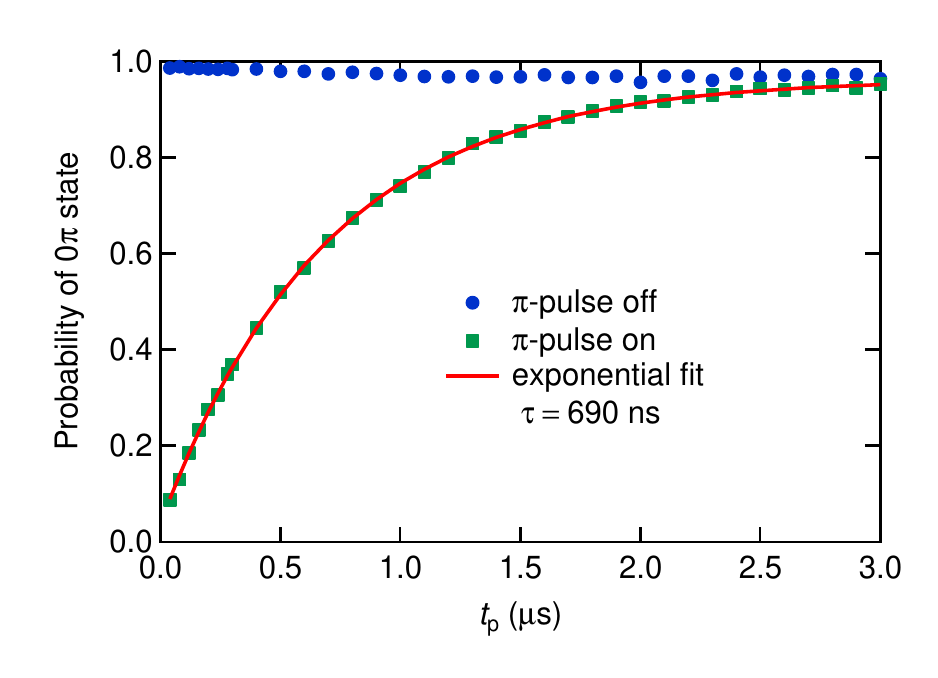}
\end{center}
\caption{Probability of $0\pi$-state as a function of the delay time $t_{\rm p}$ 
between the readout and pump pulses 
with qubit-control $\pi$-pulse on (green square) and off (blue circle). 
The pulse sequence is shown in Fig.~4a in the main article. 
Here, $t_{\rm c}=10$~ns, $t_{\rm r}=50$~ns, and $t_{\rm d}=300$~ns. 
The solid curve is an exponential fit with a time constant of 690 ns.}
\label{FigS_T1}
\end{figure}

\begin{figure}
\begin{center}
\includegraphics[width=0.56\columnwidth,clip]{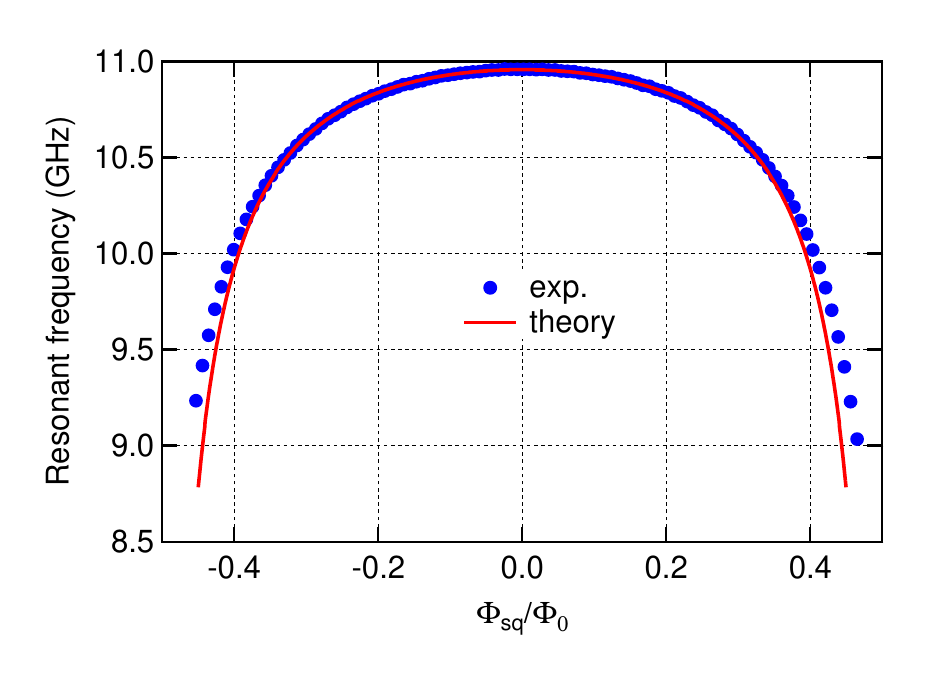}
\end{center}
\caption{Resonant frequency $\omega_0^{\rm PO}$ 
as a function of the flux bias $\Phi_{\rm sq}$. 
Solid circles represent the experimental data, and the solid curve is a 
theoretical fit (Eq.~\ref{fdep_omega0}) to the data 
for $|\Phi_{\rm sq}/\Phi_{0}|\leq 0.32$. 
In the fitting, we assumed $L_{\rm cav}$ of 1.08~nH, $C_{\rm in}^{\rm PO}$ of 15~fF, and 
$C_{\rm J}$ of 50~fF, which are designed values. 
The deviation of the fitting outside the above range could be 
due to the loop inductance of the SQUID, which is neglected in the theory and 
estimated to be $\sim 40$~pH by the simulation. 
}
\label{FigS_fr_flux}
\end{figure}

\begin{figure}
\begin{center}
\includegraphics[width=0.80\columnwidth,clip]{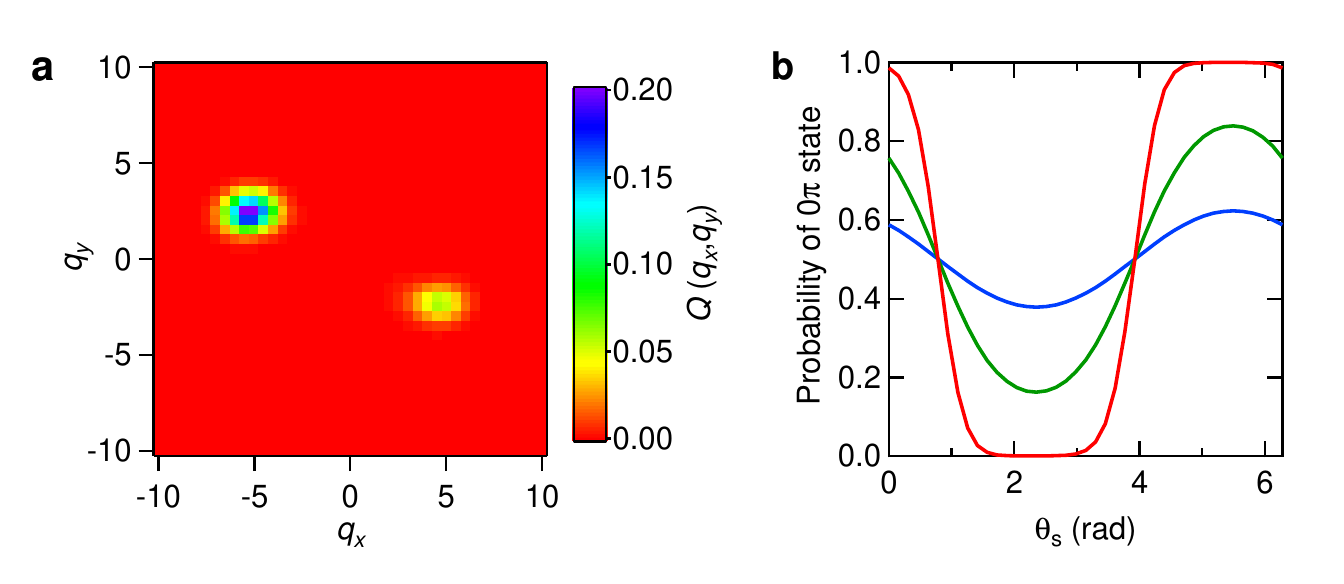}
\end{center}
\caption{Results of the numerical calculation. 
{\bf a,} Example of the calculated $Q$ function. 
The parameters used in the calculation are 
$P_{\rm p}/P_{\rm p0}=1.660$ (2.2~dB), $N^{\rm PO}=0.09$, $\theta_{\rm s}=\pi/2$. 
The density matrix is truncated at $N=80$, 
and the calculation is stopped at $\tau=20$.
{\bf b,} Probability of $0\pi$ state as a function of 
the locking signal phase $\theta_{\rm s}$ for 
$N^{\rm PO}=0.009$ (blue), 
$N^{\rm PO}=0.09$ (green), and $N^{\rm PO}=0.9$ (red). 
Other parameters used in the calculation are the same as in {\bf a}. 
}
\label{fig_th}
\end{figure}